\documentclass[a4paper,10pt]{article}
\usepackage{rotating}
\usepackage{graphicx}
\usepackage{amsmath,amstext,amssymb}
\usepackage[margin=1in]{geometry}
\def\be{\begin{equation}}
\def\ee{\end{equation}}
\def\ba{\begin{eqnarray}}
\def\ea{\end{eqnarray}}
\newcommand{\psidag}{\psi^{\dag}}

\newcommand{\chidag}{\chi^{\dag}}

\newcommand{\tildepsidag}{\tilde{\psi}^{\dag}}

\newcommand{\tildechidag}{\tilde{\chi}^{\dag}}

\begin{document}

\begin{center}
{\Large\bf Pomeron-Odderon Interactions: A Functional RG Flow Analysis}
\vskip 10pt

\large
Gian Paolo Vacca\footnote{e-mail address: vacca@bo.infn.it}
 \\
\vskip 10pt

INFN, Sezione di Bologna, via Irnerio 46, I-40126 Bologna, Italy

\vskip 20pt
{\bf Abstract}
\end{center}
In the quest for an effective field theory which could help to understand some non perturbative feature of the QCD in the Regge limit,
we consider a Reggeon Field Theory (RFT) for both Pomeron and Odderon interactions and perform an analysis of the critical theory using functional renormalization group techniques, unveiling a novel symmetry structure.

\section{INTRODUCTION}
Strong interactions in the Regge limit, well before QCD was established as a QFT fundamental description, were historically described within Regge theory,
which is still applied to analysing non perturbatively the scattering processes at Tevatron, ISR, HERA, RHIC and LHC.
The inclusion of interactions among reggeons was later encoded in the so called reggeon field theory~\cite{Gribov:1968fg,Gribov:1968uy,Abarbanel:1973pq,Sugar:1974td,Migdal:1973gz,Moshe:1977fe}.
On the other hand perturbative QCD analysis have confirmed that QCD has a special behavior when moving to the Regge kinematics, large rapidity $Y$, even if one cannot really probe the limit of large transverse distances $|x_\perp|$ (impact parameter), which requires a full non perturbative understanding of QCD.
Many perturbative results have been obtained in the framework of the so called BFKL approach~\cite{Lipatov:1976zz,Fadin:1975cb,Kuraev:1976ge,Kuraev:1977fs,Balitsky:1978ic}, based on the reggeization of the gluon. Basic features are encoded in the kernel for rapidity propagating BFKL Pomeron. Later the  Odderon~\cite{Bartels:1999yt,Bartels:2013yga} exchange was also obtained as solution of the BKP equations~\cite{Bartels:1980pe, Kwiecinski:1980wb,Bartels:2012sw} and used to give phenomenological predictions~\cite{Bartels:2001hw}. Another ingredient is given by the interaction vertices~\cite{Bartels:1994jj, Braun:1997nu, Bartels:1999aw,Bartels:2002au, Bartels:2004ef, Bartels:2004hb} which are naturally describing their interactions local in rapidity. Resummed tree level effective interactions in nuclei were also described in the dipole picture by the BK equation~\cite{Balitsky:1995ub,Kovchegov:1999yj}. Similarly results  for the Odderon have been obtained also in the Color Glass Condensate, dipole and Wilson line approaches\cite{Kovchegov:2003dm, Hatta:2005as, Kovner:2005qj}.
Impact factors\cite{Fadin:1999de,Fadin:1999df} for external partons or photons~\cite{Bartels:2002uz,Balitsky:2012bs} and vertices for jet production~\cite{Bartels:2001ge, Bartels:2002yj} have also been constructed. Unitarity constraints at next-to-leading-log (strong bootstrap) we derived~\cite{Braun:1998zj,Braun:1999uz,Fadin:2002hz}. All these results can be interpreted as showing that the QCD dynamics, in the Regge limit and at small transverse distances, tends to organize in an effective $2+1$ QFT and it is  natural to ask what happens when moving to the non perturbative region. Here we discuss recent efforts to test a local reggeon field theory as a possible candidate for such a theory.
At this stage we limit ourselves to study some non perturbative features in the critical and universal properties of some RFT in order to see if they may be compatible with QCD. Understanding, as an effective field theory, the transition from QCD to RFT is a much harder work, mainly because of essential non perturbative effects in the gauge theory.
A first analysis~\cite{Bartels:2015gou} was recently carried on for a RFT in the pure Pomeron sector, giving very interesting results. Since theoretical analysis in QCD show the presence of the Odderon composite states and interactions, a larger RFT can also be considered~\cite{Bartels:2016ecw}.

\section{A Pomeron-Odderon RFT}
We start from the fact that Pomeron and Odderon reggeons are the leading Regge poles in the "energy" plane dual to the rapidity variable which contributes to the scattering amplitudes. One takes the approximation of linear trajectories, characterized by an intercept and slope. In perturbative QCD it is well known that the Pomeron is not a pole but a cut and it is in a supercritical regime (intercept $\alpha(0)>1$, while the Odderon appears to have an intercept exactly at one (even in NLL, at least for large $N_c$ or for $N=4$ SYM)~\cite{Bartels:1999yt,Bartels:2013yga}, and the transition to a pole description is expected to be induced by constraints from the non perturbative region. Apart from the kinetic part of the reggeon action we shall then consider approximate interactions, homogeneous and of non derivative nature (ultra local) in the transverse space, i.e. described by a local potential. And finally we shall allow for an anomalous scaling of the Pomeron and Odderon fields.

In our notation $\psi,\psidag$ denote the Pomeron fields, and for the Odderon we introduce
the fields $\chi,\chidag$. The effective action has the form: 
\ba
\!\!\!\!\!\!\Gamma[\psidag,\psi,\chidag,\chi]\!\!\!\!\!&=&\!\!\!\!\!\!\!\int \!\! \mathrm{d}^2 x \,  \mathrm{d} \tau
\! \left( \!Z_{P }(\frac{1}{2} \psidag {\partial}_{\! \tau}^\leftrightarrow \psi -\alpha'_{P } \psidag\nabla^{2}\psi)
+ Z_{O }(\frac{1}{2}\chidag{\partial}_{\! \tau}^\leftrightarrow \chi -\alpha'_{O } \chidag\nabla^{2}\chi)
+ V_k[\psi,\psidag,\chi,\chidag] \!\right)
\ea
In the lowest possible truncation the local potential is characterized by just cubic interactions and is given by
\be
V_3=-\mu_P \psidag \psi +i\lambda \psidag (\psi +\psidag) \psi  -\mu_O \chidag \chi +i\lambda_2 \chidag (\psi +\psidag) \chi + \lambda_3 
(\psidag \chi^2 + {\chidag}^2 \psi).
\ee
We shall consider, systematically, interactions of much higher order in order to improve the accuracy of our results.

The structure of the potential and the allowed interactions are constrained by signature conservation (even for the Pomeron and odd for the Odderon) and by the overall sign of the multi reggeon discontinuity amplitudes $-i \prod_j (i \xi_j)$, where $\xi_j$ are the signature factors $\xi=(\tau-e^{-i\pi \omega})/\sin{\pi\omega}$ with $\omega=\alpha(0)-1$: for the Pomeron $\xi_P$ is almost imaginary while for the Odderon $\xi_O$ is almost real.
Therefore t-channel states with odd and even number of Odderons never mix. Moreover transitions $P\to PP$ are imaginary (two Pomeron cut is negative), $P\to O O$ are real (two Odderon cut is positive),  $O\to O P$ is imaginary (Odderon-Pomeron cut is negative). These considerations are easily generalized to all orders~\cite{Bartels:2016ecw}.
Essentially we can write the potential in terms of different contributions  with operators which relate states differing by an integer number of Odderon pairs.
In perturbative QCD the $P\to O O$ vertex has been computed in the generalized leading-log approximation. We shall see that the reggeon interaction nevertheless are dominated by a fixed point in which operators changing the number of Odderon pairs are not present, these being present only in deformations of the critical theory.

Next we introduce dimensionless variables. The field variables as well as the potential are rescaled as follows:
\be
\tilde{\psi} = Z_P^{1/2} k^{-D/2} \psi, \,\, \,\, \tilde{\chi}= Z_O^{1/2} k^{-D/2} \chi, \,\, \,\, \tilde{V}=\frac{V}{\alpha'_P k^{D+2}}\ ,
\label{dimless}
\ee
where $D$ is the transverse space dimension. We also need to introduce the dimensionless ratio $r=\frac{\alpha'_O}{\alpha'_P}$.

We study the critical properties of this reggeon filed theory using wilsonian functional renormalization group techniques. In particular we shall study the flow of the infrared (IR) regulated effective action, which generates the 1PI vertices, the so called effective average action.
Its exact flow with respect to the RG time $t=\log{k/k_0}$, where $k$ is the IR scale, is given by~\cite{Wetterich:1992yh,Morris:1994ie}
 \be
\partial_{t}\Gamma=\frac{1}{2}{\rm Tr}[\Gamma^{(2)}+\mathbb{R}]^{-1}
\partial_{t}\mathbb{R}\ ,
\label{eq:exactflow1}
\ee
where $\mathbb{R}$ is the infrared regulator of the two point function. The details can be found in~\cite{Bartels:2016ecw}.
The computation of the trace is typically performed passing to a Fourier representation with transverse momenta and reggeon "energy" (dual to rapidity). Therefore the Regge limit is completely understood in terms of an IR limit in the "energy"-transverse space.

In order to study the critical behavior of the theory we need to consider the flow of the dimensionless couplings, and look for the zeros of their beta functions.
Therefore in the flow of the rescaled potential one has also to consider the rescaling of the fields and of the potential as in equation~(\ref{dimless}), which gives the additional contributions to the equation defining $\partial_t \tilde{V}$, to be summed to the one of equation~(\ref{eq:exactflow1}),
\be
\left(-(D+2) + \zeta_P\right) \tilde{V} + \left(\frac{D}{2} + \frac{\eta_P}{2} \right)
\left( \tilde{\psi} \frac{\partial \tilde{V}}{\partial \tilde{\psi}} + \tildepsidag \frac{\partial \tilde{V}}{\partial \tildepsidag} \right)
+ \left( \frac{D}{2} + \frac{\eta_O}{2} \right) \left(\tilde{\chi} \frac{\partial \tilde{V}}{\partial \tilde{\chi}} + \tildechidag \frac{\partial \tilde{V}}{\partial \tildechidag} \right).
\ee
To extract the flow equations of a chosen set of couplings in the potential, which includes also the intercept, as usual one projects the flow equation on the basis of the operators considered, which in our specific case in a finite dimensional set of monomials in $\tilde{\psi},\tilde{\psi}^\dagger,\tilde{\chi},\tilde{\chi}^\dagger$. The quantum trace in equation~(\ref{eq:exactflow1}) can be evaluated at constant fields.

Then we need to determine the flow of $Z_P$, $Z_O$, $\alpha'_P$ and $\alpha'_O$. 
These informations are extracted in the one loop approximation studying the flow of the $2$-point function $\Gamma^{(1,1)}$ for both the Pomeron and the Odderon, setting the fields to be constant and looking at the dependence in the external energy $\omega$ and the invariant $q^2$ of the external transverse momentum.

\begin{figure}[h]
\centerline{\includegraphics[width=6cm]{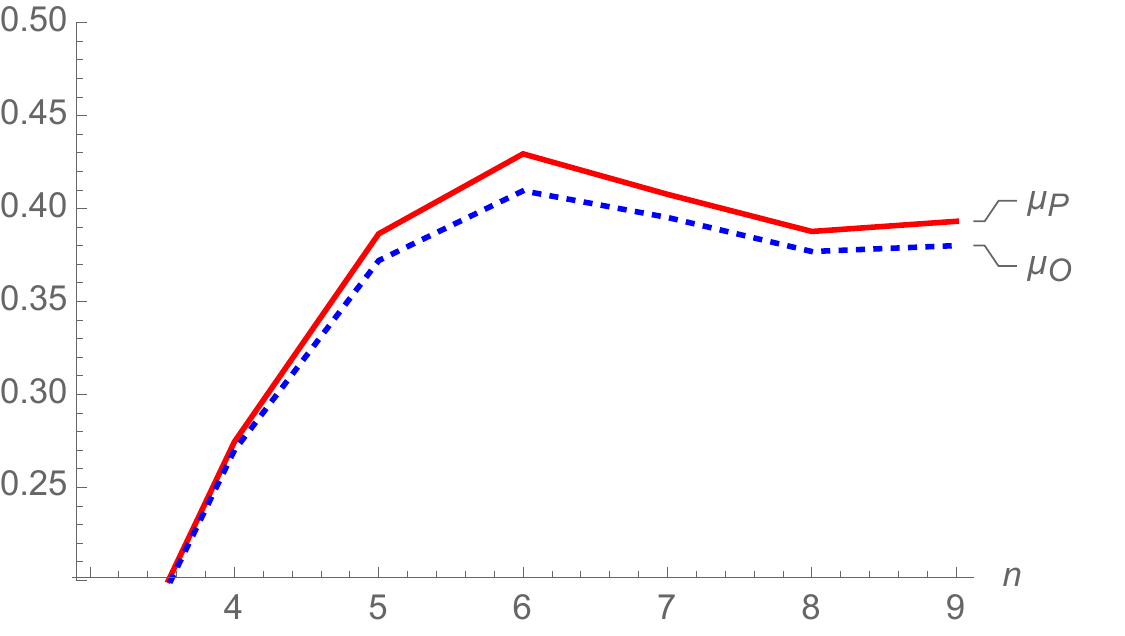}
\includegraphics[width=6cm]{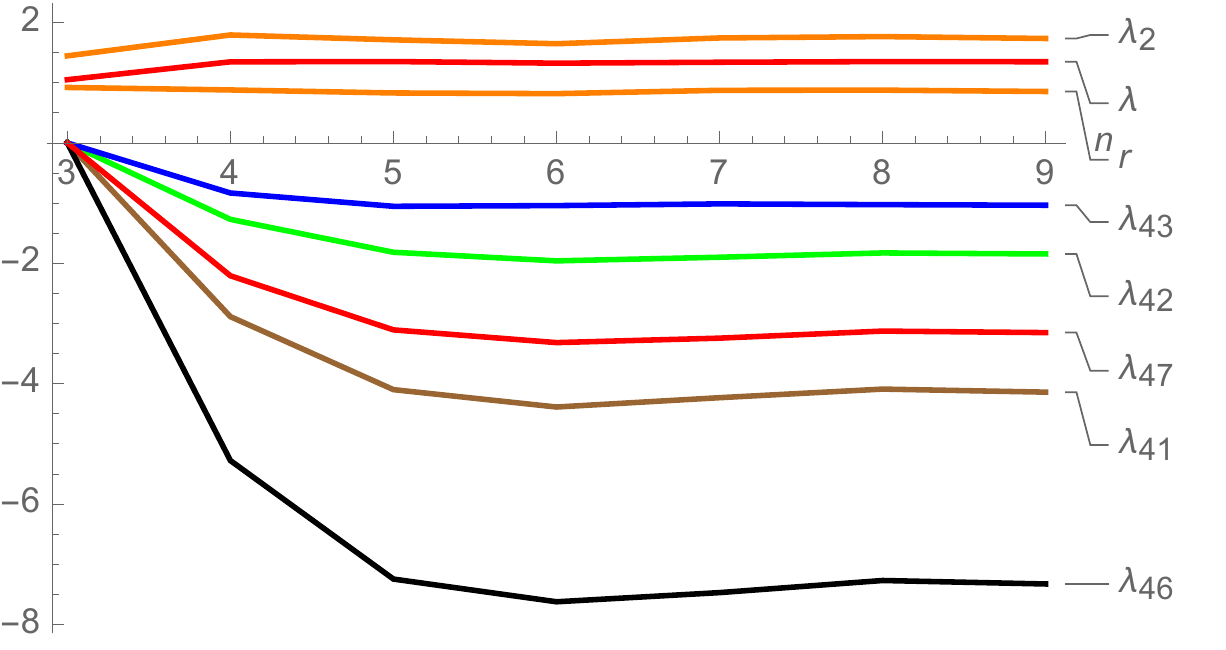}}
\caption{Values of the parameters of the fixed point solution of the LPA truncations for different orders $n$ of the polynomial ($3\le n \le 9$). 
The masses (which equal intercept minus one)  $\mu_P$ (red curve) and $\mu_O$ (blue dotted curve) for the Pomeron and Odderon fields are in the left panel.
The first non zero couplings $\lambda, \lambda_2,  \lambda_{41}, \lambda_{42}, \lambda_{43}, \lambda_{46}, \lambda_{47}, r$ are reported on the right panel.}
\label{fig_trunc}
\end{figure}

\begin{figure}[h]
\centerline{\includegraphics[width=6cm]{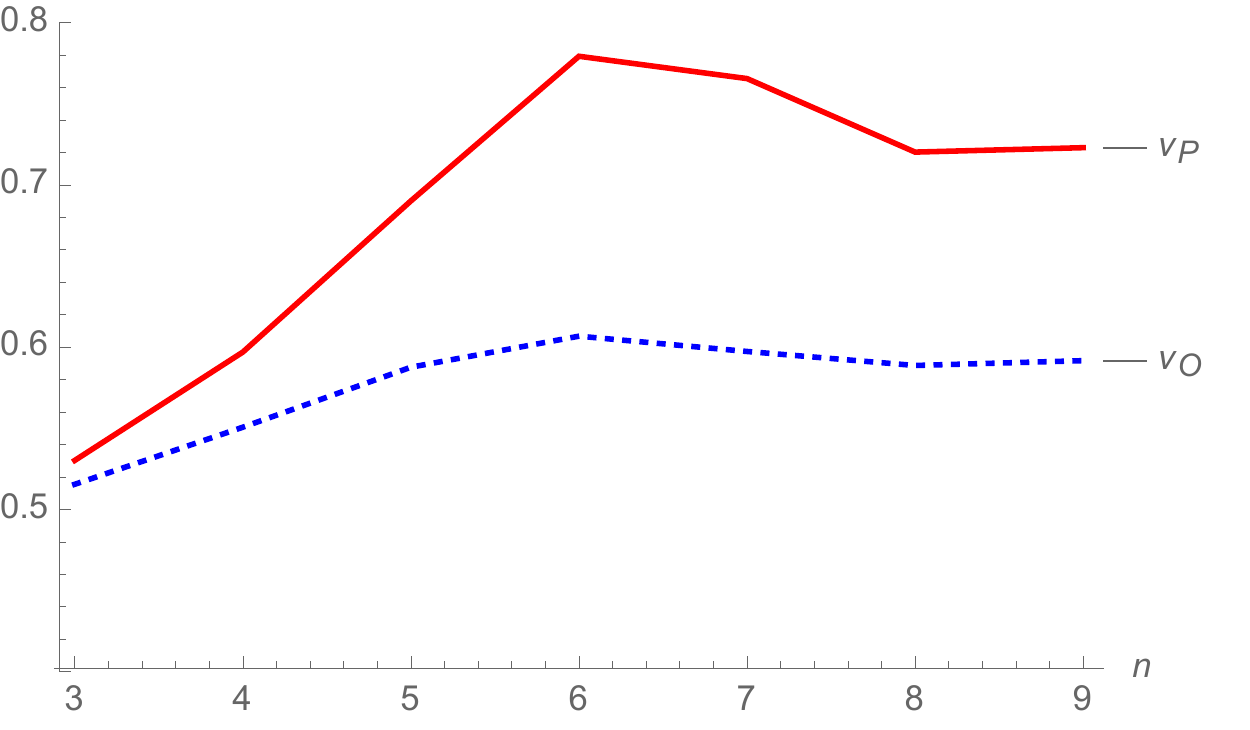}
\includegraphics[width=6cm]{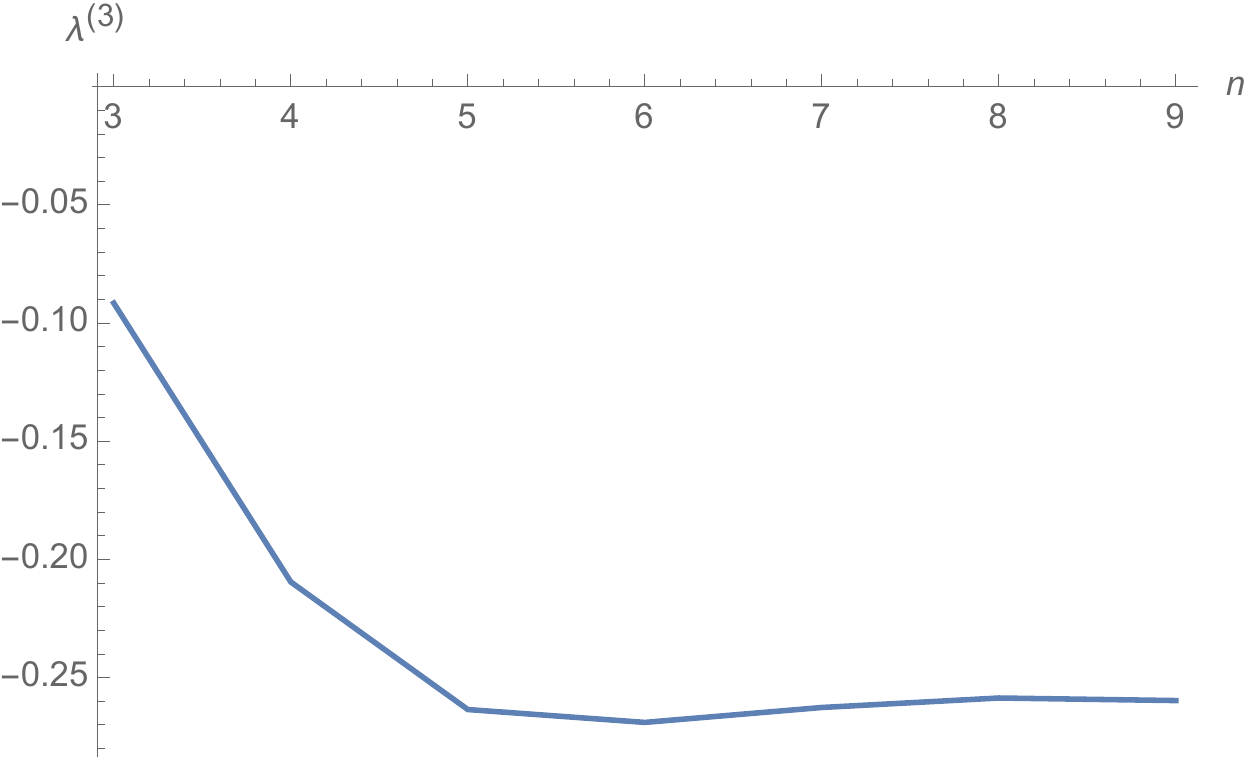}}
\caption{Values of the critical exponents of the fixed point solution of the LPA truncations for different orders $n$ of the polynomial ($3\le n \le 9$). The two negative leading eigenvalues define the two critical exponents $\nu_P$ (red curve) and $\nu_O$ (blue dotted curve) for the Pomeron and Odderon fields (left panel). We report also the value of a third negative eigenvalue found in our approximation (right panel).}
\label{fig_crit}
\end{figure}

Using this framework it is also possible to perform a one-loop $\epsilon$-expansion analysis in $D=4-\epsilon$ around the critical dimension $4$ of the transverse space. Using the cubic truncation one finds, besides a fixed point solution related to the pure Pomeron theory, another fixed point which is non trivial in the Odderon sector, such that $\lambda^2, \lambda_2^2,  \lambda_3^2, \mu_P, \mu_O=O(\epsilon)$:
\ba
\label{specialFPepsilon}
&{}&\!\!\!\!\mu_P=\frac{\epsilon}{12},\quad \lambda^2=\frac{8\pi^2}{3} \epsilon, \quad \eta_P=-\frac{\epsilon}{6}, \quad \zeta_P=\zeta_O=\frac{\epsilon}{12}, \\
&{}&\!\!\!\!\mu_O=\frac{95\!+\!17\sqrt{33}}{2304}\epsilon, \quad \lambda_2^2=\frac{23\sqrt{6}\!+\!11\sqrt{22}}{48}\epsilon, \quad \lambda_3=0,
 \quad \eta_O=-\frac{7\!+\!\sqrt{33}}{72}\epsilon,
\quad r=\frac{3}{16}(\sqrt{33}\!-\!1).\nonumber
\ea
One immediately notes that $\lambda_3=0$ so that the critical theory has no operator changing the number of Odderon pairs.
Apart from the anomalous dimensions, the universal critical exponents are extracted with a spectral analysis of the stability matrix,
obtained by a linear analysis around the fixed point.
We find two negative eigenvalues, associated to two relevant directions, and the corresponding critical exponents:
\be
\lambda^{(1)}=-2+\frac{\epsilon}{4} \,\,\,\rightarrow \nu_P=\frac{1}{2} +\frac{\epsilon}{16} \quad , \quad
\lambda^{(2)}=-2+\frac{\epsilon}{12} \rightarrow \nu_O=\frac{1}{2} +\frac{\epsilon}{48}.
\label{nuepsilon}
\ee
We than perform an analysis in the physical $D=2$ transverse space looking (numerically) at the fixed point and its spectral properties using polynomial truncations for the potential with increasing orders up to order $9$. 
We find again that there exists only a non trivial scaling solution in both Odderon and Pomeron sectors, which is characterized by the absence of operators which change the number of Odderon pairs.
We also compute the critical exponents of such a critical theory. We show the results of the analysis with two figures. In {\bf FIGURE}~\ref{fig_trunc}. we show the couplings for the potential and the convergence of the procedure. In {\bf FIGURE}~\ref{fig_crit}. we present the critical exponents and again their dependence with the order of the truncation. For more details we refer to~\cite{Bartels:2016ecw}.
Finally for the anomalous dimensions at criticality we find $\eta_P=-\partial_t \log{Z_P}\simeq -0.33$,  $\eta_O=-\partial_t \log{Z_O}\simeq -0.35$ and $\zeta_P =-\partial_t \log{\alpha'_P}= \zeta_O=-\partial_t \log{\alpha'_O}\simeq+0.17$. We expect values for the anomalous dimensions to be about $20\%$ larger in magnitude according to what we observe from Monte Carlo analysis in the pure Pomeron sector.
We find numerically three relevant directions, but one with an eigenvalue pretty close to zero which could change sign with a more accurate analysis and then turn into an irrelevant direction.
The relevant directions define the critical surface and the fate of the flow towards the IR is determined, depending on where at some scale the theory lies. If the theory lies on the critical surface then the flow points into the fixed point. Otherwise it will be repulsed along the relevant directions. In any case the dimensionful theory will be characterized by finite dimensionful couplings in the IR. Therefore depending on the "UV" RFT one may flow towards a subcritical, critical or supercritical RFT. It is not yet clear if QCD poses constraints on such possible scenarios. The theories at criticality or flowing very close to it will be characterized by the absence or suppression of Odderon number changing interactions. As discussed in~\cite{Bartels:2016ecw} this may have phenomenological consequences on how much processed involving the Odderon with high mass diffraction~\cite{ Bartels:2003zu} might be suppressed. Future studies will be devoted to understanding the QCD-RFT transition.


\end{document}